\documentclass[aps,prl,twocolumn,showpacs,preprintnumbers,amsmath,amssymb,amsfonts]{revtex4}

\usepackage{graphicx}
\usepackage{dcolumn}
\usepackage{bm}

\begin{document}
\title{Detection of synchronised chaos in coupled map networks using
symbolic dynamics}
\author{Sarika Jalan} \email[Electronic address: ]{jalan@mis.mpg.de} 
\author{Fatihcan M. Atay} \email[Electronic address: ]{atay@member.ams.org}
\author{J\"urgen Jost}\email[Electronic address: ]{jjost@mis.mpg.de}
\altaffiliation[Also at: ]{Santa Fe Institute, Santa Fe, NM 87501, USA.}
\affiliation{Max Planck Institute for Mathematics in the%
Sciences, Leipzig 04103, Germany}

\date{\today}

\begin{abstract}
We present a method based on symbolic dynamics for the detection
of synchronization in networks of coupled maps and distinguishing between
chaotic and random iterations. The symbolic dynamics are defined using
special partitions of the phase space which prevent the occurrence of
certain symbol sequences related to the characteristics of the dynamics.
Synchrony in a
large network can be detected using measurements from only a single node by
checking for the presence of forbidden symbol transitions. The method
utilises a relatively short time series of measurements and hence is
computationally very fast. Furthermore, it is robust against parameter
uncertainties, is independent of the network size, and does not require
knowledge of the connection structure.
\end{abstract}

\maketitle

Symbolic dynamics is a fundamental tool available to describe complicated
time evolution of a chaotic dynamical system, the Smale horseshoe being a
most famous prototype \cite{smale}. In symbolic dynamics, instead of
representing a trajectory by infinite sequences of numbers one watches the
alternation of symbols. In doing so one 'loses' some information but certain
important invariants and robust properties of the dynamics may be kept \cite%
{book-sym2}. Most studies of symbolic dynamics in the literature are
based on the so-called generating partition \cite{generating-book} of the 
phase space, for which topological entropy of the system achieves its 
supremum \cite%
{topological-entropy}. On the other hand, finding a generating partition 
for general systems is a challenging open problem and is the issue of the
constant research \cite{generating-eg}. The
misplacement of partition leads to 'diminishing' of the computed entropy of
the system \cite{mispart}. Symbolic dynamics based on generating partitions
plays a crucial role in understanding many different properties of dynamical
systems. However, as we shall show, dynamics based on 
certain other partitions also have important practical uses. 
In the following, we construct
specific partitions such that certain symbolic transitions do not appear in
the time evolution of a dynamical system. We present two applications, namely 
distinguishing
between chaotic and random iterations and the detection of synchronization
in networks. In the first application, the absence of these forbidden
transitions in a chaotic iteration distinguishes it from a corresponding
random iteration. In the second application, we utilise the fact that in a
synchronized network all nodes follow the same evolution as an isolated
unit, so that synchrony can be detected by comparing the symbolic dynamics
from any single node against the dynamics in isolation.

Synchronisation in large ensembles of coupled dynamical units is 
studied in many different fields \cite{syn-eg,syn-book,syn-motor,%
Epilepsy-book,Mormann03,syn-comm}. There are several systems
where synchronisation processes are significant, e.g.,
information processing in the brain \cite{syn-motor}.
The detection of synchronization of an
extended system from local measurements has important applications. For
instance, certain pathologies in the neural system, such as epileptic
seizures, manifest themselves by synchronized brain signals 
\cite{Epilepsy-book} and there is some evidence that they can be predicted by
changing levels of synchronization \cite{Mormann03}. 
It is thus of interest
to be able to determine different levels of synchronization in a brain area
from local measurements, such as EEG recordings. Similarly, fast and easy
detection of synchronization in communication or power networks is
important for efficient management of such large systems \cite{syn-comm}.
The methods so far known to detect synchronisation are based on
the decrease in the topological entropy or on the correlation between the
bivariate or multivariate time series observed at different nodes
using tools from nonlinear time
series analysis \cite{book-tseries1,tseries-syn3}. The first 
method
requires the construction of phase space from time series using delay
coordinates \cite{book-tseries1,tseries-syn1}, which itself is not
a trivial task and typically requires extensive computational effort 
\cite{embedding}. Furthermore, its success depends on the availability of a 
sufficiently long time series, although it is not clear what length is 
sufficient \cite{embed}. Even with an infinitely long time series, the 
detection of synchronisation
relies on the decrease of phase space dimension, and it is not clear whether
any decrease is due to the synchronisation of nodes or to
the unsynchronised low-dimensional chaos. Similarly, distinguishing chaotic and
random systems is not always easy, and the issue is of both fundamental and
practical importance \cite{ran-chaos}. In this Letter, we shall address
these problems using symbolic dynamics based on specific partitions.

\begin{figure}
\includegraphics[height=3.1cm,width=5cm]{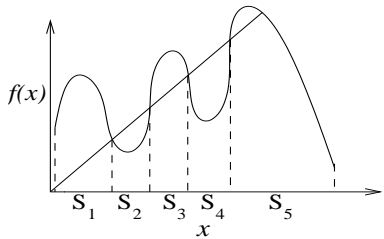}
\caption{A partition showing avoiding sets. In the pairs $(S_2,S_3)$,%
$(S_2,S_4)$, $(S_2,S_5)$, $(S_4,S_5)$, $(S_3,S_1)$, $(S_3,S_2)$, $(S_5,S_1)$, 
$(S_5,S_2)$, $(S_5,S_3)$, $(S_5,S_4)$, $(S_5,S_5)$, the first set avoids
the second.}
\label{part-1d}
\end{figure}

We consider dynamical systems defined by the iteration rule
\begin{equation}
x(t+1)=f(x(t))  \label{system}
\end{equation}%
Where $t\in {\mathbb Z}$ is the discrete time and $f:S\rightarrow S$ is a map
on a subset $S$ of ${\mathbb R}^{n}$. Let $S_{i}:i=1,\ldots ,m$ be a partition
of $S$, i.e., a collection of mutually disjoint subsets satisfying $\cup
_{i=1}^{m}S_{i}=S$. We assume that the $S_{i}$ are nonempty and $m\geq 2$ to
prevent trivial cases. The symbolic dynamics corresponding to (\ref{system})
is the sequence of symbols $\{{\dots ,s_{t-1},s_{t},s_{t+1},\dots \}}$ where 
$s_{t}=i$ if $x(t)\in S_{i}$. We say the set $S_{i}$ {\em avoids }$S_{j}$ if 
\begin{equation}
f(S_{i})\cap S_{j}=\emptyset .  \label{av}
\end{equation}%
Clearly, if $S_{i}$ avoids $S_{j}$, so does any of its subsets. We also talk
about a {\em self-avoiding set} if (\ref{av}) holds with $i=j$. The essence
of our method is based on choosing a partition where (\ref{av}) holds for
one or more sets in the partition. The significance is that if $S_{i}$
avoids $S_{j}$, then the symbolic dynamics cannot contain the symbol
sequence $ij$. The notion is extended in a straightforward way to the $k$th
iterate of $f$. Thus, if $f^{k}(S_{i})\cap S_{j}=\emptyset $, then the
symbol sequence for the dynamics cannot contain the symbols $i$ and $j$ at
two positions which are $k-1$ symbols apart. This constrains the symbol
sequences that can be generated by a given map, and provides a robust method
to distinguish between different systems by inspecting their symbolic
dynamics. As examples of avoiding sets, we mention that for the common
unimodal maps of the interval $[0,1]$, such as the tent or logistic maps,
the set $(x^{\ast },1]$ and its subsets are self-avoiding, where $x^{\ast }$
denotes the positive fixed point of $f$. More generally,
if $f$ is a continuous one-dimensional function on the interval $[r_1,r_2]$ having
$m$ fixed points $x_1^{\ast}, x_2^{\ast},..
x_m^{\ast}$, and the derivatives $f^{'}(x^*)$ exist, then 
one can define a partition with $S_1=[r_1,x^*_1]$,
$S_i =(x^{\ast}_{i-1},x^{\ast}_{i}]$ for $i=2,...,m$,
and $S_{m+1}=(x_m^{\ast},r_2]$. Then one can have several avoiding pairs, 
namely (Figure \ref{part-1d}),
\begin{eqnarray}
\text{if }f^{\prime }(x_{i}^{\ast })&>&0, \text{ then }f(S_{i})\cap S_{j}=\emptyset
,\forall i<j  \nonumber \\
\text{if }f^{\prime }(x_{i}^{\ast }) &<&0, \text{ then }f(S_{i})\cap S_{j}=\emptyset
,\forall i>j. \nonumber 
\end{eqnarray}%
Before turning to their applications, we first show that partitions which
contain avoiding sets always exist and give a constructive proof.
Suppose one starts with some partition of $m$ sets for which (\ref{av}%
) does not hold for any $i,j$. Now fix some pair $i\neq j$. If $%
f(S_{i})\subset S_{j}$, then $S_{i}$ can be arbitrarily partitioned into two
parts which avoid each other. If, on the other hand, $f(S_{i})\not\subset
S_{j}$, then let $S_{m+1}=f^{-1}(S_{j})\cap S_{i},$ and re-define $S_{i}$ as 
$S_{i}^{\prime }=S_{i}\backslash S_{m+1}$ to obtain a new partition with $%
m+1 $ sets, where $S_{i}^{\prime }$ avoids $S_{j}$. The same argument can be
used to construct self-avoiding sets: Assume $f(S_{i})\not\subset S_{i}$
(otherwise further partition $S_{i}$ to obtain a set which is not invariant
under $f$)\footnote{%
This is possible except for the trivial case when $f$ is the identity
map.} and define $S_{m+1}=f^{-1}(S_{i})\cap S_{i}$ and $S_{i}^{\prime
}=S_{i}\backslash S_{m+1},$ so that $S_{i}^{\prime }$ is self-avoiding.
Hence, it is possible to modify a given partition so that the sequence $ij$
never occurs in the symbolic dynamics.

We first consider the problem of distinguishing 
chaotic and random iterations. To be specific, suppose the iteration
rule is given by the tent map%
\begin{equation}
f(x)=(1-2|x-\frac{1}{2}|).  \label{tent-map}
\end{equation}%
Its stationary density on [0,1] is the uniform density, $p(x)\equiv 1$. If
we take the generating partition, which is simply a partition of $[0,1]$ at
the middle, then the above tent map and a random sequence from a uniform
distribution give equal Kolmogorov-Sinai (KS) entropy \cite%
{KS-tseries}. This can be seen by considering the permutation
entropy based on transition probabilities of length $d,$ 
\begin{equation}
PE=-\frac{1}{d}\sum_{i_{1},i_{2}...}^{m}p(i_{1},i_{2}...i_{d})\ln
(p(i_{1},i_{2},....,i_{d})  
\end{equation}%
where $p(i_{1},i_{2},\dots ,i_{d})$ is the joint probability that $x(1)\in
S_{i_{1}}$, $x(2)\in S_{i_{2}}$, $\dots $, $x(d)\in S_{i_{d}}$. Taking $d=2$
and $m=2$, the iteration takes a trajectory to the left or
right partition with equal probabilities, and the situation is identical to
the tossing of a fair coin. Hence, for this case the symbolic dynamics does
not distinguish between the random and chaotic sequences, and the
permutation entropy for both systems is equal to $\ln 2$. Now we shift the
partition by $\delta $, i.e take the separation point at $l_{p}=1/2+\delta $%
, and calculate the entropies of the tent map and the random iteration with
the same probability distribution, as a function of $\delta $.It turns out
that the difference between the two entropies is maximum when $l_{p}$ is
chosen as the fixed point $a/(a+1)$ of the tent map. For the corresponding
partition, i.e, $S_{1}=[0,l_{p}],S_{2}=(l_{p},1]$, the transition $%
2\rightarrow 2$ does not occur. So, the random sequence can be identified by
the presence of this transition. As the partition point is moved further to
the right in the range $a/(a+1)\leq l_{p}<1$, the transition remains
forbidden, but the difference between entropies decreases, reaching zero at $%
l_{p}=1$. The implication is that a longer time series would now be needed
as the observed occurrence of the symbol $2$ becomes less frequent. Hence,
the optimal partition would be the one for which the self-avoiding set $%
S_{2} $ is largest, which also yields the largest entropy difference between
the chaotic and random maps.
Note that the choice of $l_p$ in the range $(1/2, a/(a+1))$ gives a non-generating
partition, but has no two-symbol forbidden sequences. Although 
some longer sequences may be forbidden, their detection requires 
longer time series and more computational effort. On the other hand, 
avoiding sets, which forbid short sequences, is faster and can utilize 
shorter time-series.
\begin{figure*}
\includegraphics[width=14cm,height=7.2cm]{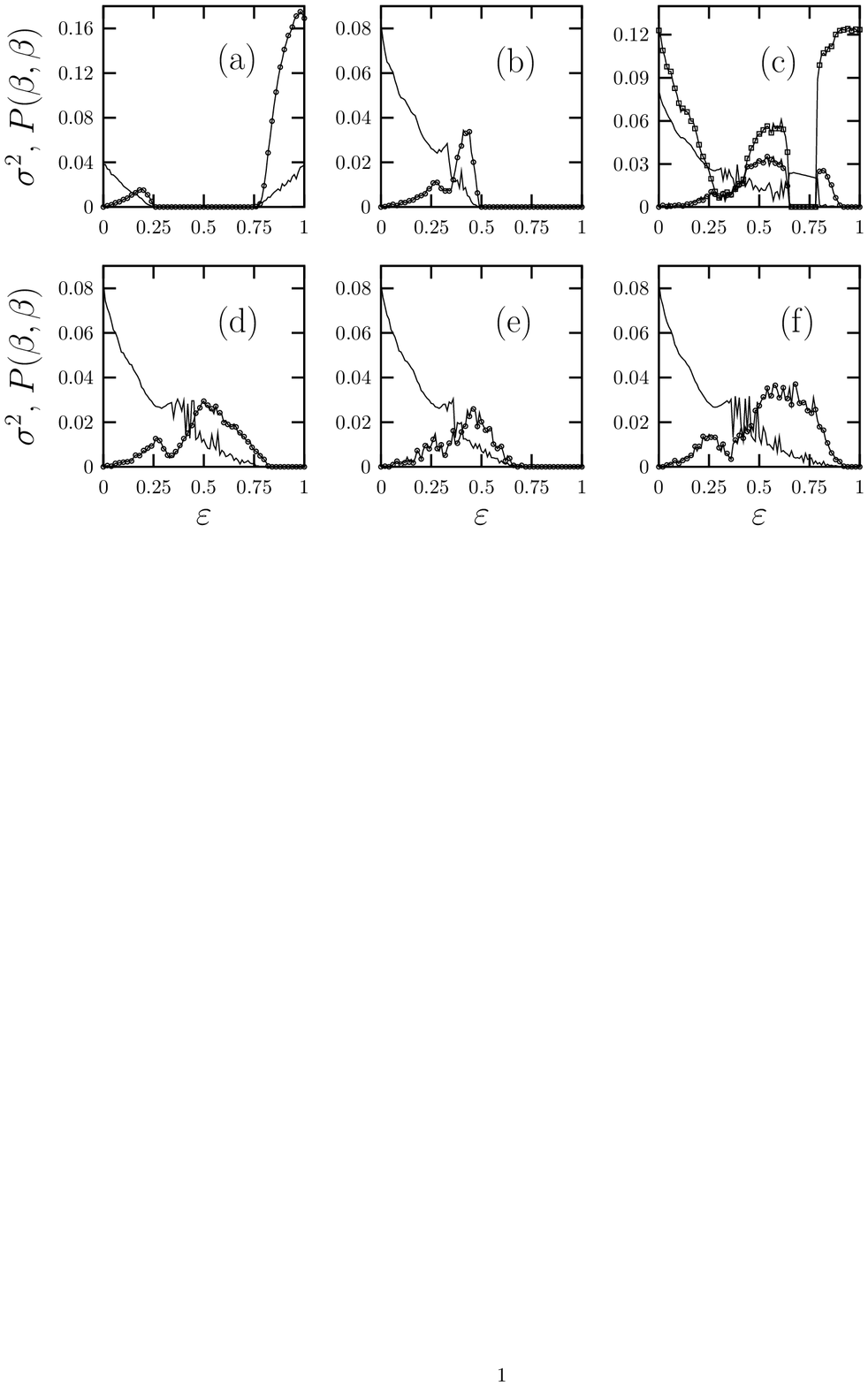}
\caption{Detection of synchronisation in various networks of coupled maps.
Figures (a)-(d) are plotted for coupled tent maps (\ref{tent-map})
and (e)-(f) are for coupled logistic maps ($4 x (1-x)$). The networks consist 
of (a) 
two coupled maps, (b) a globally coupled network of size 50, (c) a small-world
network of size 100 and average degree 30, (d) a scale-free network of size 100
and average degree 15 constructed by the algorithm given in \cite{BA}, 
(e) a scale-free network of size 200 and average degree 10, and (f) a 
random network of size 200 and average degree 8. The
horizontal axis is the coupling strength, and the vertical axis gives the
synchronization measure $\protect\sigma^2$ (--) for the whole network, 
as well as the transition
probability $P(\beta,\beta)$ ($\circ$) calculated using a scalar time series from a
randomly selected node. In all cases, the synchronization region ($\protect%
\sigma^2 = 0$) coincides with the region where $P(\beta,beta)=0$ and the other
transition probabilities are nonzero, which is the situation for the
uncoupled map. Note that in subfigure (c) there is an interval of $\protect%
\epsilon$ roughly between $0.65$ and $0.8$ such that $P(\beta,\beta)=0$, but there
is no synchronization as $P(\alpha,\alpha)$ (shown by $\square$) is also zero here, 
unlike the case for the isolated tent map.}
\label{sym-sigma-tent}
\end{figure*}
We next turn to the detection of synchronization in networks. Consider the
well-known coupled map lattice model%
\begin{equation}
x_{i}(t+1)=f(x_{i}(t))+\frac{\varepsilon }{k_{i}}\sum_{j=1}^{N}C_{ij}\left[
f(x_{j}(t))-f(x_{i}(t))\right]  \label{cml}
\end{equation}%
where $x_{i}(t)$ is the state of the $i$th node at time $t$, $i=1,\dots ,N$, 
$C_{ij}$ are the elements of the adjacency matrix with value 1 or 0
depending upon whether $i$ and $j$ are connected or not, $\varepsilon \in
\lbrack 0,1]$ is the coupling strength, and $k_{i}$ is the number of
neighbours of the $i$th unit. Clearly, the system (\ref{cml}) can exhibit a
much wider range of behaviour than (\ref{system}) depending on the coupling
structure and the choice of parameters; so the corresponding symbol
sequences observed from a node can vary widely. However, at the synchronized
state $x_{i}(t)=x_{j}(t)$ for all $i,j$ and $t$, all
nodes evolve according to the rule (\ref{system}). It follows that when the
network is synchronized, the symbolic sequence measured from a node will be
subject to the same constraints as that generated by (\ref{system}). If the
partition used to define the symbolic dynamics contains avoiding sets, one
has a simple way of detecting synchronization of the network by simply
choosing a random node and observing the presence or absence of symbol
subsequences which are forbidden in (\ref{system}).

For numerical illustration, we take $f$ to be the tent map (\ref{tent-map})
with the partition $S_{1}=[0,x^{\ast }],$ $S_{2}=(x^{\ast },1],$ at the
fixed point $x^{\ast }=a/(a+1)$, so that the transition $2 \to 2$ has zero probability 
for the isolated map.
In actual calculations, we use a slightly modified way 
to define the symbolic dynamics: To
two consecutive measurements $x_{t}x_{t+1}$ we assign the symbol $\alpha$ if $%
x_{t+1}\geq x_{t}$ and the symbol $\beta$ otherwise \cite{note}. This is completely
equivalent to the symbol sequences defined by the sets $S_{1},S_{2}$ because
of the specific partition point $x^{\ast }$; thus the transition $%
\beta\rightarrow \beta$ does not occur for the single tent map. The advantage of
using $\alpha,\beta$ is that one only needs to check increases and decreases in the
measured signal, which adds more robustness in case the fixed point is not
precisely known. We evolve (\ref{cml}) starting from random initial
conditions and estimate the transition probabilities using time series of
length $\tau =1000$ from a randomly selected node. Note that the length of
the time series is much shorter time than would be required by standard
time-series methods which use embedding to reconstruct the phase space for
large networks \cite{embed}. In our case, however, the length of the series
is independent of the network size. We estimate the transition probability $%
P(i,j)$ by the ratio $\sum_{t}n(s_{t}=i,s_{t+1}=j)/\sum_{t}n(s_{t}=i)$,
where $n$ is a count of the number of times of occurrence. Synchronisation
is signalled when the variance of variables over the network given by $%
\sigma ^{2}=\left\langle \frac{1}{N-1}\sum_{i}[x_{i}(t)-\bar{x}%
(t)]^{2}\right\rangle _{t}$ drops to zero, where $\bar{x}(t)=\frac{1}{N}%
\sum_{i}x_{i}(t)$ denotes an average over the nodes of the network and $%
\left\langle \dots \right\rangle _{t}$ denotes an average over time. Fig.~%
\ref{sym-sigma-tent} summarizes the results. It is seen in all cases that
the region for synchronization exactly coincides with the range for which $%
P(\beta,\beta)$ is zero (and other transition probabilities are nonzero; see
subfigure (c)). Hence, regardless of network topology and size, both
synchronized and unsynchronized behavior of the network can be accurately
detected over the whole range of coupling strengths using only measurements
from an arbitrarily selected node.

We note that the method applies equally well to any diffusive network, say%
\begin{equation}
x_{i}(t+1)=f(x_{i}(t))+\sum_{j=1}^{N}w_{ij}g(x_{i}(t),x_{j}(t))
\label{network}
\end{equation}%
which may be asymmetrically coupled or weighted, with weights given by $%
w_{ij}$, $x$ is $n$-dimensional, and $g$ is a diffusion function satisfying $%
g(x,x)=0$ $\forall x$. The basic idea is again the same: So long as the
synchronized dynamics is identical to the isolated dynamics, synchrony can
be detected by comparing the transition probabilities of symbolic sequences
to those for the isolated map. The fact that certain transition
probabilities are exactly zero makes this procedure especially robust, and
the choice of the partition plays an important role here.
For networks including transmission delays,
the synchronized dynamics is no longer identical 
to that of the isolated map \cite{delay}.
Nevertheless, it can be shown that the properties of the map  
still induces certain forbidden transitions, whose presence or absence can 
be used to detect synchrony in the delayed network.
Due to space constraints, we defer these results to an extended paper.

In conclusion, we have presented a simple and effective method based on
symbolic dynamics for the detection of synchronisation in diffusively
coupled networks. The method works by taking measurements from as few as one
single node, and can utilize rather short sequences of measurements, and
hence is computationally very fast. Furthermore, it is robust against noise
and parameter uncertainties, is independent of network size, and does not
require knowledge of the connection structure. Additionally, the method
allows to distinguish between chaotic and random iterations with same
underlying probability density.


\begin{thebibliography}{10}
\bibitem{note} Such a comparison between successive values of the
variables has been introduced in \cite{PE} for the calculation of  
permutation entropy, and used in \cite{SJ1} 
for studying phase synchronisation
in coupled map networks. The first study \cite{PE} has
shown that permutation entropy based on the comparison of the n-successive
values gives a good estimate of the KS entropy, and if $n$ is large enough, it
completely matches the KS entropy.
\end{thebibliography}
\end{document}